\documentclass[reprint,prd,11pt,onecolumn,notitlepage]{revtex4-1}
\usepackage[english]{babel}
\usepackage{amsmath,amssymb,graphicx,bm,lipsum,float}
\usepackage[colorlinks=true,citecolor=blue,linkcolor=blue,urlcolor=blue]{hyperref} 
\usepackage[font={small},flushleft,indent]{caption}
\usepackage{setspace}
\usepackage{subcaption}
\usepackage{placeins}
\usepackage{overpic}

\begin{document}    
\title{Flux Enhancement of Corotating Hotspots in Accretion Disks from the Slow-Light Effect}
\author{Qing-Hua Zhu}  
\email{zhuqh@cqu.edu.cn} 
\affiliation{Department of Physics and Chongqing Key Laboratory for Strongly Coupled Physics, Chongqing University, Chongqing 401331, China}

\begin{abstract}

With the Event Horizon Telescope and future Very Long Baseline Interferometry arrays poised to image supermassive black holes, there is a strong motivation to understand the dynamic aspects of the accretion flow near the black hole.  Interestingly, in such highly relativistic regime, the finite light-travel time should be taken into account to correctly simulate the images, known as the slow-light effect.
This paper investigates the impact of the slow-light effect on the observational signatures of hotspots corotating with the Keplerian disks.  It is found that the magnification can be modulated by the hotspot's orbital velociy. Specifically, the corotating hotspots at the maxima of magnification are located at the image positions shifted relative to the positions behind the black hole. This subsequently causes the peak of the magnification profile to shift towards alignment with  the peak of the redshift factor profile, ultimately leading to flux enhancement relative to the results of previous studies. This enhancement becomes particularly pronounced for corotating hotspots in the strong-field regime at large inclination angle, indicating that the slow-light effect is essential for accurately modeling high-energy emission in the vicinity of the black hole.
\end{abstract} 
\maketitle

\section{Introduction\label{I}} 
 
With the Event Horizon Telescope \cite{EventHorizonTelescope:2019dse,EventHorizonTelescope:2022wkp}, GRAVITY \cite{GRAVITY:2018sef,GRAVITY:2023avo}, RadioAstron \cite{Valtonen:2025ovs,Kim:2023hnr}, and future ground- and space-based Very Long Baseline Interferometry (VLBI) arrays \cite{Johnson:2023ynn,Johnson:2024ttr} poised to image diverse populations of supermassive black holes, renewed interest is expected in exploring the black hole spacetime geometry \cite{Kumar:2018ple,Shaikh:2019hbm,Khodadi:2020jij,EventHorizonTelescope:2020qrl,Wielgus:2021peu,EventHorizonTelescope:2021dqv,Vagnozzi:2022moj,EventHorizonTelescope:2022xqj} and accreting matter \cite{Narayan:2019imo,Chael:2018gzl,Ressler:2020voz,Ripperda:2021zpn,EventHorizonTelescope:2021srq,Narayan:2021qfw,EventHorizonTelescope:2022urf} in the strong field regime. In additional to static imaging of a black hole, recent studies have also explored the potential by leveraging time-dependent phenomena, such as the studies on intensity variability from accretion disk dynamics \cite{Hadar:2020fda,EventHorizonTelescope:2022exc,EventHorizonTelescope:2022ago,EventHorizonTelescope:2022okn,Conroy:2023kec,Cardenas-Avendano:2024sgy,Zhu:2025jqh,Zhang:2025vyx} or gravitational fluctuations \cite{Wang:2019skw,Zhu:2023omf,Zhong:2024ysg}, binary black hole systems \cite{Davelaar:2021eoi,Davelaar:2021gxx,Wang:2025bmh}, and orbiting flare on accretion disks \cite{Matsumoto:2020wul,Antonopoulou:2024qco,Yfantis:2023wsp,Yfantis:2024eab,GRAVITY:2020lpa,Xie:2025skg}.

For flares on the accretion disk, the emission region is localized to a small portion of the disk and is modeled as a hotspot \cite{1992A257594B,1994ApJ...425...63B,Broderick:2005jj,Broderick:2005my,Li:2014fza,Ripperda:2020bpz,Rosa:2022toh,Genzel:2024vou}. The temporal radiation flux at specific frequency from the hotspot, also known as its light curve, is given by
\begin{eqnarray}
	F & = & \int g^3 I_{\text{emt}} \textrm{d} \Omega = \int g^3 I_{\text{emt}}
	\Sigma \textrm{d} \sigma~,~ \text{ with  }\Sigma  \equiv  {\textrm{d} \Omega}/{\textrm{d} \sigma}~, \label{1}
\end{eqnarray}
where $g$ is redshift factor, $I_{\text{emt}}$ is surface emission intensity of the source, $\textrm{d} \Omega$ is solid angle of the images on observer's celestial sphere, and $\textrm{d} \sigma$ is spatial surface of the source.
The flux of a hotspot is governed by three distinct factors: a) the intrinsic emission mechanism, denoted by $I_{\text{emt}}\mathrm{d}\sigma$; b) relativistic redshift shifts $g$, including Doppler redshift and gravitational redshift; and c) image distortion due to gravitational light bending, quantified by $\Sigma$. The first factor involves the local physics of the sources, whereas the latter two are of relativistic effects. In this study, we mainly focus on the image distortion, formulated by $\Sigma$, of corotating hotspot distributed on the accretion disk. 

Within the standard framework of gravitational lensing, the magnification factor is a well-defined quantity used to describe image distortion \cite{Schneider:1992bmb,Perlick:2004tq,Saha:2024axf}, thereby providing an indication for underlying spacetime geometry \cite{Bozza:2002zj,Islam:2020xmy,Furtado:2020puz,Atamurotov:2021hoq,Ali:2021psk,Fu:2021fxn,Kuang:2022ojj,Kumar:2022fqo,Junior:2023xgl,Xie:2024dpi,Lan:2025hlv,Cheong:2025lwp}. 
While the critical role of $\Sigma$ in characterizing the magnification effect for the hotspots in the strong-field regime has been well established \cite{Li:2014fza,Tian:2019yhn,Boero:2026dup,Frost:2025wvh}, formulating a magnification factor for sources at finite distances remains non-trivial \cite{Perlick:2003vg,Bozza:2010xqn,Frost:2025wvh}. Specifically, the difficulty arises because the source position in curved spacetime can not be determined via a straight line connecting the source and observer. 
For imaging a black hole in the near region, this situation can be alleviated given that the accretion flows near black holes are modeled in the coordinates defined by the spacetime metric, such as the general relativistic magnetohydrodynamics (GRMHD) simulations \cite{DeVilliers:2002ab,Gammie:2003rj,Chael:2018gzl,Ressler:2020voz, Ripperda:2020bpz,Ripperda:2021zpn}. In this framework, the surface element $\mathrm{d}\sigma$ on the accretion disk corresponds to the grid cells output from GRMHD simulations. Therefore, there could be no need to concern with how to correctly formulate the source position in the coordinate system via `visual' definition in the absence of gravity \cite{Schneider:1992bmb,Saha:2024axf} or a local frame of sources \cite{Zhu:2025jqh,Frost:2025wvh}.  
In this sense, the magnification effect of a source in strong-field regime could be studied without conceptual ambiguity.


When exploring the fine structure of imaging of time-variable sources, the slow-light effect, which takes into account the finite light-travel time in curved spacetime to produce images, has been demonstrated to have a non-negligible impact on the variable emission such as the accretion disks \cite{Dexter:2010pe,Bronzwaer:2018lde,Moscibrodzka:2021fxv,White:2022paq} and the relativistic jets \cite{Jeter:2018fyo,MacDonald:2021lkt,Saiz-Perez:2024jkh,Tsunetoe:2025nla}. 
It is also equally crucial to account for the slow-light effect in the modeling of infrared flares, which are interpreted as bright hotspots on relativistic orbits around the black hole \cite{GRAVITY:2018sef,GRAVITY:2023avo}.
However, the previous analytical studies on the radiation flux from the orbiting hotspot appear not to fully account for the slow-light effect \cite{1992A257594B,1994ApJ...425...63B,Zhu:2024vxw}. 
Thus, one of the objectives of this study is to identify the effects neglected in these studies.

In this paper, we investigate the hotspot corotating with Keplerian accretion disks, incorporating the slow-light effect. 
Central to our approach is the recognition that quantify $\Sigma$ in Eq.~(\ref{1}) should be modulated by the slow-light effect.  To isolate the pure slow-light contribution, we introduce a relative magnification factor through decomposing $\Sigma$. We present the distribution of the relative magnification factor across the accretion disk and its temporal variability. Based on the behavior of the magnification factor, we examine whether the radiation flux from orbiting hotspot is enhanced due to the slow-light effect.


The remainder of this paper is organized as follows. In Sec.~\ref{II}, we briefly review the ray-tracing method and celestial coordinates adopted in this study. In Sec.~\ref{III}, we introduce the relative magnification factor and validate the method with static disk around a black hole. In Sec.~\ref{IV}, we perform a time-series reconstruction  framework to realize the slow-light effect, and study properties of the relative magnification factor distributed on accretion disks. 
In Sec.~\ref{V}, we examine the light curve of the corotating hotspot, and demonstrate its flux enhancement due to the slow-light effect. 
In Sec.~\ref{VI}, the conclusions and discussions are summarized.

\section{Ray-tracing method and celestial coordinates \label{II}}

We briefly review the ray-tracing method and the corresponding imaging on the observer's celestial sphere. It is implemented for a Schwarzschild black hole. The metric is given by 
\begin{eqnarray}
	\textrm{d} s^2 & = & - f (r) \textrm{d} t^2 + \frac{\textrm{d} r^2}{f (r)} + r^2 (\textrm{d}
	\theta^2 + \sin^2 \theta \textrm{d} \phi^2)~,
\end{eqnarray}
where $f (r) = 1 - 2 M / r$. A light ray emitted at $\bm x_\text{s}[ =(r_\text{s}, \theta_\text{s}, \phi_\text{s})]$ and received by the observer at $\bm x_\text{o}[=(r_\text{o}, \theta_\text{o}, \phi_\text{o})]$ can be described by the solution of geodesic equations, namely, \cite{Zhu:2024vxw}
\begin{subequations}
	\begin{align}
	0 & =  \rho I_r (r_\text{o}, r_\text{s} ; \rho) - G_r (\theta_\text{o}, \theta_\text{s} ; \varphi)~,\label{4a}\\
	t_\text{o} - t_\text{s} & =  I_t (r_\text{o}, r_\text{s} ; \rho)~, \label{4b}\\
	\phi_\text{o} - \phi_\text{s} & =  - \sin \theta_\text{o} \cos \varphi G_{\phi} (\theta_\text{o}, \theta_\text{s} ; \varphi)~,  \label{4c}
\end{align} \label{4}
\end{subequations}
where the constants $\rho$ and $\varphi$ can be derived from the impact parameters of light rays, and
\begin{subequations}
	\begin{align}
	I_r & =  \pm_r \int_{r_\text{s}}^{r_\text{o}} \textrm{d} r \left\{ \frac{1}{r \sqrt{r^2 -  \rho^2 f}} \right\}~,\\
	G_r & =  \pm_{\theta} \int_{\theta_\text{s}}^{\theta_\text{o}} \textrm{d} \theta \left\{  \frac{1}{\sqrt{1 - \csc^2 \theta \sin^2 \theta_\text{o} \cos^2 \varphi}} \right\}~,\\
	I_t & =  \pm_r \int_{r_\text{s}}^{r_\text{o}} \textrm{d} r \left\{ \frac{r}{f \sqrt{r^2 -  \rho^2 f}} \right\}~,\\
	G_{\phi} & =  \pm_{\theta} \int_{\theta_\text{s}}^{\theta_\text{o}} \textrm{d} \theta \left\{  \frac{1}{\sin^2 \theta \sqrt{1 - \csc^2 \theta \sin^2 \theta_\text{o} \cos^2  \varphi}} \right\}~.
\end{align}
\end{subequations}
Given the locations of the source $\bm x_\text{s}$ and the observer $\bm x_\text{o}$, one can determine the impact parameters of the light rays using Eqs.~(\ref{4a}) and (\ref{4c}), defined as the functions $\rho(\bm x_\text{s},\bm x_\text{o})$ and $\varphi(\bm x_\text{s},\bm x_\text{o})$  \cite{Zhu:2024vxw}. Physically, this implies that an observer at  $\bm x_\text{o}$ receives a light ray specified by  $(\rho,\varphi)$ emitted by the source at $\bm x_\text{s}$.

The celestial coordinates determine the location of the received light rays on the observer's sky, which can be established using several reference sources for the observer. Following the previous studies  \cite{Zhu:2024vxw,Zhu:2025jqh}, the relation between the celestial coordinate $(\Phi, \Psi)$ and the parameters $(\rho, \varphi)$ of light ray can be given by
\begin{eqnarray}
	\Phi  =  \varphi~, &~&~
	\Psi  = \arccos \sqrt{1 - \frac{\rho^2 f_\text{o}}{r_\text{o}^2}}~, \label{5}
\end{eqnarray}
where $f_\text{o}\equiv f(r_\text{o})$. The celestial coordinates $(\Phi, \Psi)$  are direct observables associated with astrometry \cite{GRAVITY:2023avo}.

\section{Relative magnification factor on accreting disk \label{III}}

The magnification effect of source near a black hole play a important role in the radiation flux of source, as described by the quantity $\Sigma$ in Eq.~(\ref{1}). Here, the $\Sigma$ is not a dimensionless quantity and relies on observer's setup, such as the observer's distance to the black hole and inclination angle relative to the accretion disk. Therefore, we introduce the relative magnification factor, a dimensionless quantity, to restrict our attention to the pure slow-light effect. 
It is important to clarify that the relative magnification factor serves purely as a mathematical tool to isolate the physical effect. This formulation must not alter the radiation flux given by Eq.~(\ref{1}). 

\subsection{Relative magnification factor\label{IIIA}}


The definition of the relative magnification factor is based on two assumptions: i) it is proportional to the ratio of the source image size defined on the observer's sky to the source size described in metric coordinates on the accretion disk; and ii) it  normalizes to unity when the point source is located in front of the black hole relative to the observer.

\begin{figure}
	\includegraphics[width=1\linewidth]{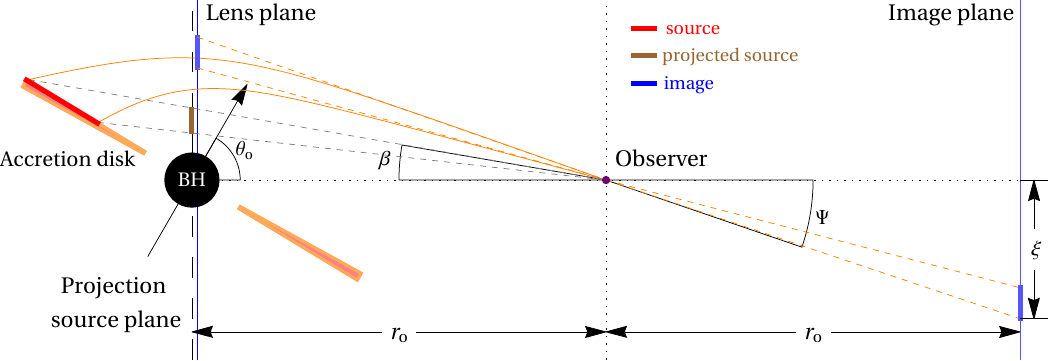}
\caption{Schematic diagram illustrating the description of the lensing system for source in the near region of the black hole. The source plane is set to be the surface of the accretion disk with inclination angle $\theta_\text{o}$ with respect to the observer. The observer locates at the distance $r_\text{o}$ from the lens plane, image plane, and also the black hole. The projected source plane and the image plane (lens plane) correspond the angular coordinate $\bm \beta$ and celestial coordinate $\Theta$, respectively. \label{F1a}}
\end{figure}

As illustrated in Fig.~\ref{F1a}, the source plane is set as the surface of the accretion disk, and the image plane (also referred to as the lens plane) is located at the distance $r_\text{o}$ from the observer. Since the observer is far away from the black hole, $r_\text{o}/M\gg1$, the images can be projected onto a plane with negligible distortion relative to the celestial sphere. We introduce a projected source plane  parallel to the image plane.  The surface element $\mathrm{d}\sigma$ on the source plane is mapped to $\mathcal{P}\mathrm{d}\sigma$ on the projected source plane, as shown by the correspondence between red and brown regions in Fig.~\ref{F1a}. Here, the $\mathcal{P}$ is the projection coefficient, relying on inclination angle of the accretion disks. 

Based on the picture in Fig.~\ref{F1a}, the relative magnification factor is defined as the Jacobian determinant relating the source-plane coordinates $\bm \beta$ to the image-plane coordinates $\Theta[\equiv(\Psi,\Phi)]$.  This factor reduces to the ratio of the size on the image plane to that on the projected source plane, namely,
\begin{eqnarray}
	\mu & \equiv &  \left|\frac{\partial\bm\Theta}{\partial \bm \beta}\right| \simeq   \frac{\xi\Delta\xi \Delta\varphi}{\mathcal{P}\Delta\sigma}
	= \frac{f_\text{o}\rho}{\mathcal{P}\sqrt{h}} \left| \frac{\partial (\rho, \varphi)}{\partial (r_\text{s}, \phi_\text{s})}
	\right|~, \label{6}
\end{eqnarray}
where $f_\ast \equiv f(r_\ast)$, $(\xi,\varphi)$ are the coordinates defined on the image plane, and we have $\sqrt{h}=\sqrt{f_\text{s}}r_\text{s}\sin\theta_\text{s}$. The $\rho$ and $\varphi$ can be derived from the impact parameters of light rays as illustrated in Sec.~\ref{II}. The third equality holds for distant observer $r_\text{o}/M \gg1$ and is derived using the $\xi\equiv r_\text{o} \Psi \simeq \rho\sqrt{f_\text{o}}$ from Eq.~(\ref{5}).   It shows that the magnification factor is proportional to the ratio of area element on the image plane to that on the source plane, satisfying the assumption i). For the thin disk at $\theta_\text{s}=\pi/2$, the projection coefficient reduces to $\mathcal{P}  \simeq  \sqrt{f_\text{s}}| \cos \theta_\text{o}|$. Here, the factor $\cos\theta_\text{o}$ was elaborated as the projection effect \cite{Tian:2019yhn} and $\sqrt{f_\text{s}}$ is an empirical calibration coefficient based on the assumption ii) to ensure $\mu=1$ for unlensed sources. In fact, the relative magnification factor calibrated by $\mathcal{P}$ is not unique. An alternative $\mathcal{P}$  is presented in Appendix~\ref{A}. 

For a spherical black hole, the Jacobi determinant in Eq.~(\ref{6}) can be derived from the geodesic equations given in Eqs.~(\ref{4}), namely,
\begin{subequations}
	\begin{eqnarray}
	\textrm{d} \varphi & = & \frac{1}{\sin \theta_\text{o}} \left( \frac{\partial}{\partial  \varphi} (\cos \varphi G_{\phi}) \right)^{- 1}  \left( \textrm{d} \phi_\text{s} - \sin \theta_\text{o} \cos \varphi \frac{\partial G_{\phi}}{\partial \theta_\text{s}} \textrm{d}
	\theta_\text{s} \right)\\
	\textrm{d} \rho & = & \left( \frac{\partial}{\partial \rho} (\rho I_r)  \right)^{- 1} \Bigg( \frac{1}{\sin \theta_\text{o}} \left( \frac{\partial}{\partial  \varphi} (\cos \varphi G_{\phi}) \right)^{- 1} \frac{\partial G_r}{\textrm{d}  \varphi} \textrm{d} \phi_\text{s} 
	\nonumber \\ &&  + \left( \frac{\partial G_r}{\partial \theta_\text{s}} -  \cos \varphi \left( \frac{\partial}{\partial \varphi} (\cos \varphi  G_{\phi}) \right)^{- 1} \frac{\partial G_r}{\partial \varphi} \frac{\partial  G_{\phi}}{\partial \theta_\text{s}} \right) \textrm{d} \theta_\text{s}   - \rho \frac{\partial
	I_r}{\partial r_\text{s}} \textrm{d} r_\text{s} \Bigg)
\end{eqnarray} \label{7}
\end{subequations} 
Using the explicit expression of $G_\phi$ (see Ref.~\cite{Zhu:2024vxw}) for the thin accretion disk $\theta_\text{s}=\pi/2$, the Jacobi determinant in Eq.~(\ref{6}) can derived from Eq.~(\ref{7}) in the form of \cite{1992A257594B,Zhu:2024vxw}
\begin{eqnarray}
	\left| \frac{\partial (\rho, \varphi)}{\partial (r_\text{s}, \phi_\text{s})} \right| & = &
	\frac{\rho}{r_\text{s} \sqrt{r_\text{s}^2 - \rho^2 f_\text{s}}} \left( \frac{\partial}{\partial
	\rho} (\rho I_r) \right)^{- 1}  \left(\frac{  1 - \sin^2 \theta_\text{o} \cos^2 \varphi }{\cos \theta_\text{o} }\right)~. \label{8}
\end{eqnarray}
This analytical result was well-established decades ago \cite{1992A257594B,1994ApJ...425...63B} (also reviewed in Ref.~\cite{Bozza:2010xqn}), and recently was extended to the regime of higher-order images \cite{Zhu:2024vxw,Zhu:2025jqh}. 

\begin{figure}
	\includegraphics[width=0.95\linewidth]{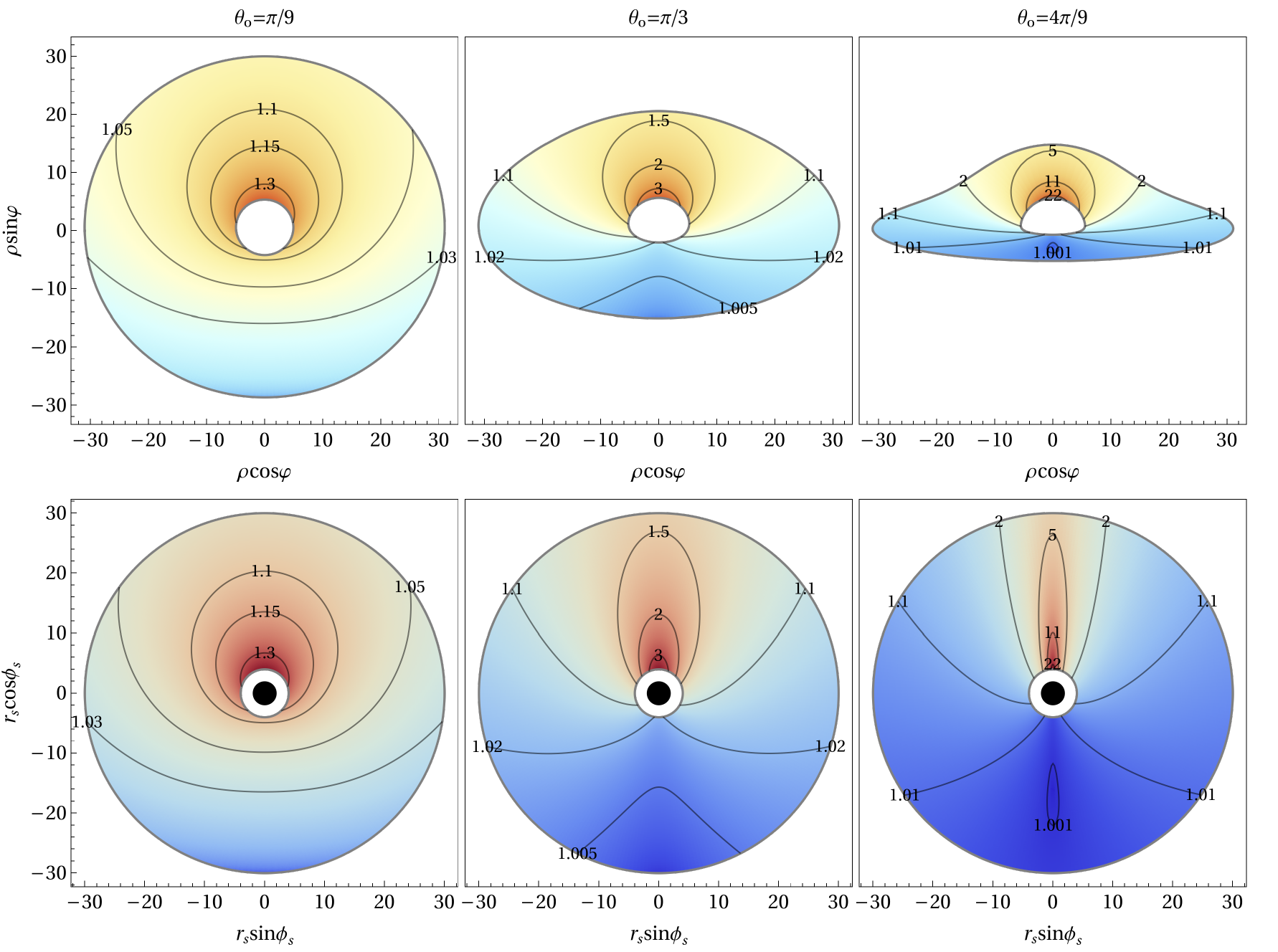} 
\caption{Relative magnification factor of point sources statically distributed on the accretion disk represented on image plane (top panels) and on source plane (bottom panels) for selected inclination angles. \label{F3}}
\end{figure}

In Fig.~\ref{F3}, we present the magnification factor of point sources distributed on the accretion disk. 
The inclination angle also contributes to an increase of the magnification factor, because point sources passing behind the black hole cause the light to undergo stronger deflection. Indeed, this behavior is well-understood within the framework of gravitational lensing: the magnification factor increases as a source approaches a caustic point \cite{Bozza:2010xqn}. For a Schwarzschild black hole, the caustic line is located   behind the black hole relative to the observer.

Recalling the expression of the projection coefficient $\mathcal{P}\simeq \sqrt{f_\text{s}}\cos\theta_\text{o}$, it depends on  both $r_\text{s}$ and $\theta_\text{o}$. 
For sources located in front of the black hole, the relative magnification factor can reduce to unity across various values of $r_\text{s}$ and $\theta_\text{o}$, as shown in Fig.~\ref{F3}. This result indicates that assumption ii) is robustly satisfied.

\subsection{From radiation flux to magnification factor}

The factor $\Sigma$ in Eq.~(\ref{1}) is the key to correctly quantify flux of the hotspot on accretion disk. However, the $\Sigma$ can not be interpreted as magnification factor $\mu$ alone, and should be decomposed into the product of three parts, namely,
\begin{eqnarray}
	\Sigma = \frac{\sin \Psi}{\sqrt{h}} \left| \frac{\partial (\Phi,
	 \Psi)}{\partial (r_\text{s}, \phi_\text{s})} \right| =:\mathcal{D}_\text{o}   \times
	 \mathcal{P} \times \mu~, \label{9}
\end{eqnarray}
where we have set $\mathrm{d}\Omega = \sin\Psi \mathrm{d}\Phi \mathrm{d} \Psi$ and $\mathrm{d}\sigma=\sqrt{h}\mathrm{d}r_\text{s}\mathrm{d}\phi_\text{s}$ in Eq.~(\ref{1}). 
And the projection coefficient $\mathcal{P}$ and magnification factor $\mu$ have been introduced in Sec.~\ref{IIIA}. 

The factor $\mathcal{D}_\text{o}$ quantifies the impact of the observer's distance from the black hole. We can derive the expression of $\mathcal{D}_\text{o}$ by making use of Eqs.~(\ref{5}), (\ref{6}) and (\ref{9}), namely,
\begin{eqnarray}
	\mathcal{D}_\text{o} & = & \mu^{-1}\mathcal{P}^{-1}\Sigma= \frac{\sin \Psi}{f_\text{o}\rho} \left| \frac{\partial (\Phi,
	\Psi)}{\partial (\rho, \varphi)} \right| = \frac{1}{r_\text{o} \sqrt{r_\text{o}^2 -
	\rho^2 f_\text{o}}}~.\label{10}
\end{eqnarray}
It shows that the $\mathcal{D}_\text{o}$ has dimensions of inverse length squared. It is derived from the ratio of the surface element on the celestial sphere to that on the image plane, reflecting the relationship between angular size and planar area.  For a distant observer, the function $f_\text{o}$ approaches unity and the angular size of the source becomes small (i.e., $\Psi \rightarrow 0$), leading to $\mathcal{D}_\text{o}\simeq r_\text{o}^{-2}$. This result is consistent with the intuition that radiation flux decays according to the inverse-square law.

Therefore, Eq.~(\ref{9}) presents the explicit relationship between the quantity $\Sigma$ for the radiation flux \cite{1992A257594B} and the relative magnification factor $\mu$. Namely, the quantity $\Sigma$ of a point source on the accretion disk is governed by the inverse-square law of flux, the projection effect, and the magnification effect.

\section{Slow-light effect for moving sources\label{IV}}

The relative magnification factor derived from Eqs.~(\ref{6}) and (\ref{8}) is limited because it neglects motion of sources on the accretion disk.
To properly formulate the magnification effect for moving sources, it is necessary to employ time-series reconstruction to ensure that the images correspond to light rays arriving at the observer simultaneously. For the static sources described in Sec.~\ref{II}, the ray tracing relies solely on Eqs.~(\ref{4a}) and (\ref{4c}). For moving sources, in contrast, it is necessary to incorporate Eq.~(\ref{4b}) to generate a correct image. In this section, we present the numerical method for simulating moving sources and explore its influences on the properties of the relative magnification factor 

\subsection{Numerical implements for the magnification factor of corotating sources\label{IVA}}

We restrict our attention to the sources corotating with the accretion disks. A set of corotating point sources $\bm x_\text{s}^{(i)}(t_\text{s})$ are given by
\begin{eqnarray}
	r_\text{s}^{(i)}= r_\text{s,0}^{(i)}~, &
	\theta_\text{s}^{(i)}= \theta_\text{s,0}~, &
	\phi_\text{s}^{(i)}= \phi_\text{s,0}^{(i)} + \omega^{(i)} t_\text{s}~. \label{15}
\end{eqnarray}
These point sources, labeled with $i$, have different values of $r_\text{s,0}^{(i)}$ and $\phi_\text{s,0}^{(i)}$, and are distributed across the thin accretion disk with $\theta_\text{s,0}=\pi/2$. For Keplerian disks, we have $\omega= \sqrt{M/(r_\text{s,0}^{(i)})^3}$. 
The kinematic evolution of the accretion disk is described by the coordinate time $t_\text{s}$ of the sources, while on the image plane, the observed time $t_\text{o}$ is adopted. The procedure, termed time-series reconstruction, generates an image at a specific time $t_\text{o}$  by integrating ray-traced sources from multiple distinct time slices $t_\text{s}$, accounting for finite light-travel time \cite{White:2022paq,Tsunetoe:2025nla}. 
This study implements the time-series reconstruction by integrating Eq.~(\ref{4b}) and Eqs.~(\ref{15}), namely,
\begin{eqnarray}
	t_\text{o} & = & \frac{\phi^{(i)}_\text{s}-\phi^{(i)}_\text{s,0}}{\omega^{(i)}}+I_t (r_\text{o}, r_\text{s}^{(i)} ; \rho(\bm x_\text{s}^{(i)}, \bm x_\text{o}))~, \label{16}
\end{eqnarray}
where the function $\rho(\bm x_\text{s}, \bm x_\text{o})$ is the main result of the ray-tracing framework described in Sec.~\ref{II} (see details in Ref.~\cite{Zhu:2024vxw}). 
For fixed observed time $t_\text{o}$ and observer position $\bm x_\text{o}$, Eq.~(\ref{16}) can be inverted to numerically determine the source coordinate $\bm x_\text{s}$, where the resulting solutions are denoted as $\bm x_{\text{s},\ast}$. The image at $t_\text{o}$ is constructed by tracing all point sources indexed by $i$, located at positions $\bm x_{\text{s},\ast}^{(i)}$. 

\begin{figure*}
	\centering
	\begin{subfigure}[t]{0.55\textwidth} 
	\includegraphics[width=1\linewidth]{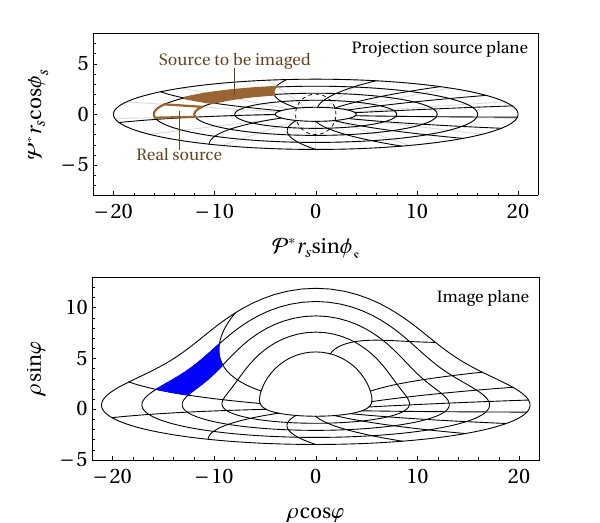}
		\end{subfigure} 
	 \begin{subfigure}[t]{0.44\textwidth} 
		\includegraphics[width=1\linewidth]{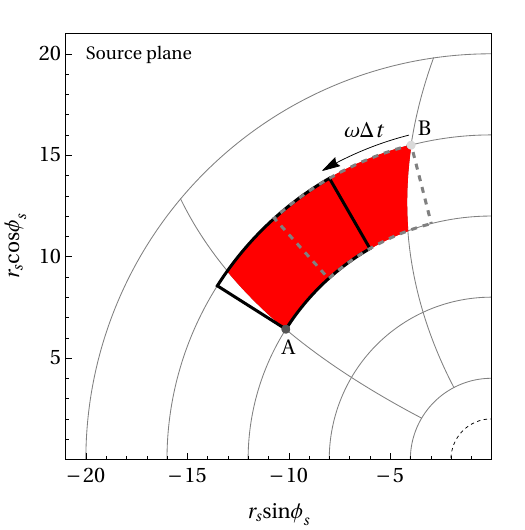}
			\end{subfigure}
			\caption{\textit{Top-left panel}: the source  (brown region) is distributed on the accretion disk in the projected source plane, denoted by $\bm x^{(i)}_\text{s}$ (real source), and $\bm x^{(i)}_{\text{s},\ast}$ (source to be imaged). \textit{Bottom-left panel}: Corresponding image of the source (blue region) distributed on the accretion disk.  \textit{Right panel}: illustration of the relationship between the  $\bm x^{(i)}_\text{s}$ and the $\bm x^{(i)}_{\text{s},\ast}$ on the source plane. Points A and B correspond to sources at different time slices $t_\text{s}^{(\text{A})}= t_\text{s}^{(\text{B})}+\Delta t$ (indicated by the region with solid and dashed boundaries, respectively), yet they are received simultaneously at observer time $t_\text{o}$. Here, all points on the red region share the same observer time $t_\text{o}$.  We set the accretion disk with a constant rotation speed of $\omega^{(i)}=(24\sqrt{3})^{-1}$ and center of the colored source region located at $r_\text{s}=14M$. \label{F4}}
\end{figure*}

We illustrate the effect of source motion on the image in Fig.~\ref{F4}. Due to the finite light-travel time, a displacement arises between the real locations of the sources and their time delayed locations, as depicted in the top-left panel of Fig.~\ref{F4}. According to the images shown in the bottom-left panel, we find that the image, i.e., the blue region, get enlarged because of the sources in motion. 
Here, the observer can see light accumulatively from multiple time slices when the sources approach the observer as shown in the right panel of Fig.~\ref{F4}.  Consequently, the magnification factor is enhanced. Conversely, not all regions on the accretion disk exhibit such enhancement. The observer could receive light rays from the region corresponding to less than one time slice when the sources recede from the observer.  In this case, the magnification factor is suppressed.  In summary, the slow-light effect induces a time delayed location and modulates the magnification factor.

For the light curves of the corotating hotspots, previous studies only included the time-delay effect, but did not include the modulation of the magnification  \cite{1992A257594B,1994ApJ...425...63B,Zhu:2024vxw}.
In order to include the latter effect, we present a rigorous calculation of the magnification factor as follows
\begin{eqnarray}
	\mu_\text{mov}  = \mu \big|_{\bm x_{\text{s}}=\bm x_{\text{s},\ast}} =   \frac{f_\text{o}\rho}{\mathcal{P}\sqrt{h(\bm x_{\text{s},\ast})}} \left| \frac{\partial (\rho, \varphi)}{\partial (r_{\text{s},\ast}, \phi_{\text{s},\ast})}
	\right|~, \label{11}
\end{eqnarray}
where the $\bm x_{\text{s},\ast}$ is the solution of Eq.~(\ref{16}) based on the time-series reconstruction. This contrasts with the simplified prescription for the magnification effect in Eq.~(\ref{6}) used in previous studies. The difference between $\mu$ and $\mu_\text{mov}$ is the main result of this study.
Since $\bm x_{\text{s},\ast}$ in Eq.~(\ref{16}) must be solved numerically, no analytical expression exists for the magnification factor in this case. In the subsequent section, we will obtain the relative magnification factor $\mu_\text{mov}$ with numerical method via separately calculating the source size on the projected source plane and the corresponding image size on the image plane, as illustrated in Fig.~\ref{F4}.

\subsection{Magnification factor distribution for corotating sources on the accretion disk}

\begin{figure}
	\includegraphics[width=1\linewidth]{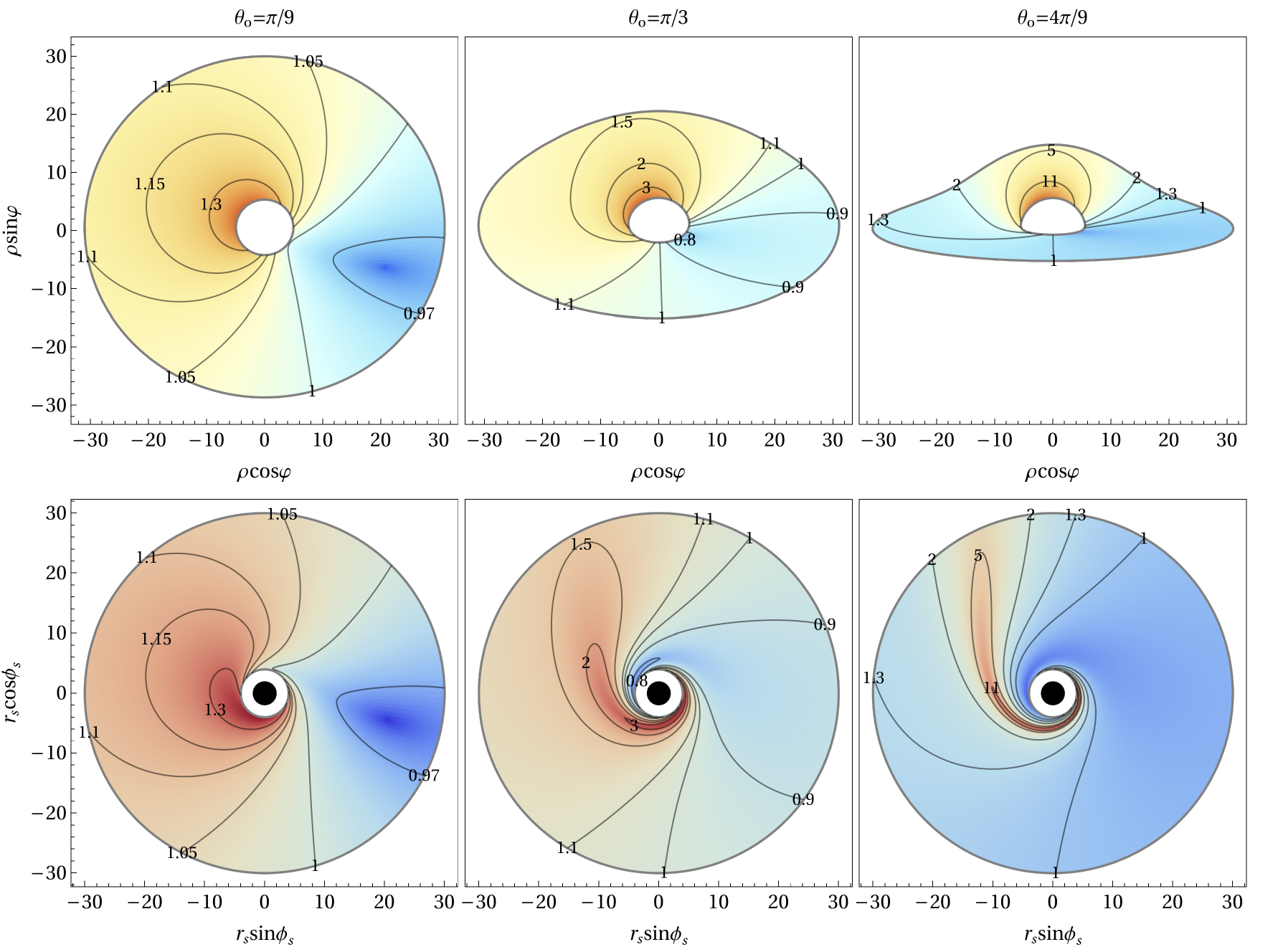}
\caption{Relative magnification factor of corotating point sources distributed on the Keplerian disk on image plane (top panels) and on source plane (bottom panels) for selected inclination angles.  \label{F5}} 
\end{figure}

Employing the time-series reconstruction in Sec.~\ref{IVA}, we present the distribution of the relative magnification factors of the point sources on the Keplerian disk [$r_\text{s,0}^{(i)}\omega^{(i)}= (M/r_\text{s,0}^{(i)})^{1/2}$] in Fig.~\ref{F5}. Because the accretion disk is in a steady state in the spacetime, the magnification factor pattern exhibits time-independent morphology across the disk. Notably, Fig.~\ref{F5} shows that the distributions of the magnification factor on both the image and source plane are significantly distorted, relative to the case for a static disk (Fig.~\ref{F3}). 
The magnification factor  is enhanced with the inclination angle, which is qualitatively consistent with the case of static disk shown in Fig.~\ref{F3}. 
The most interesting result is that the maxima of magnification factor for given $r_\text{s}$ are displaced from the region behind the black hole. 

\begin{figure}
	\includegraphics[width=1\linewidth]{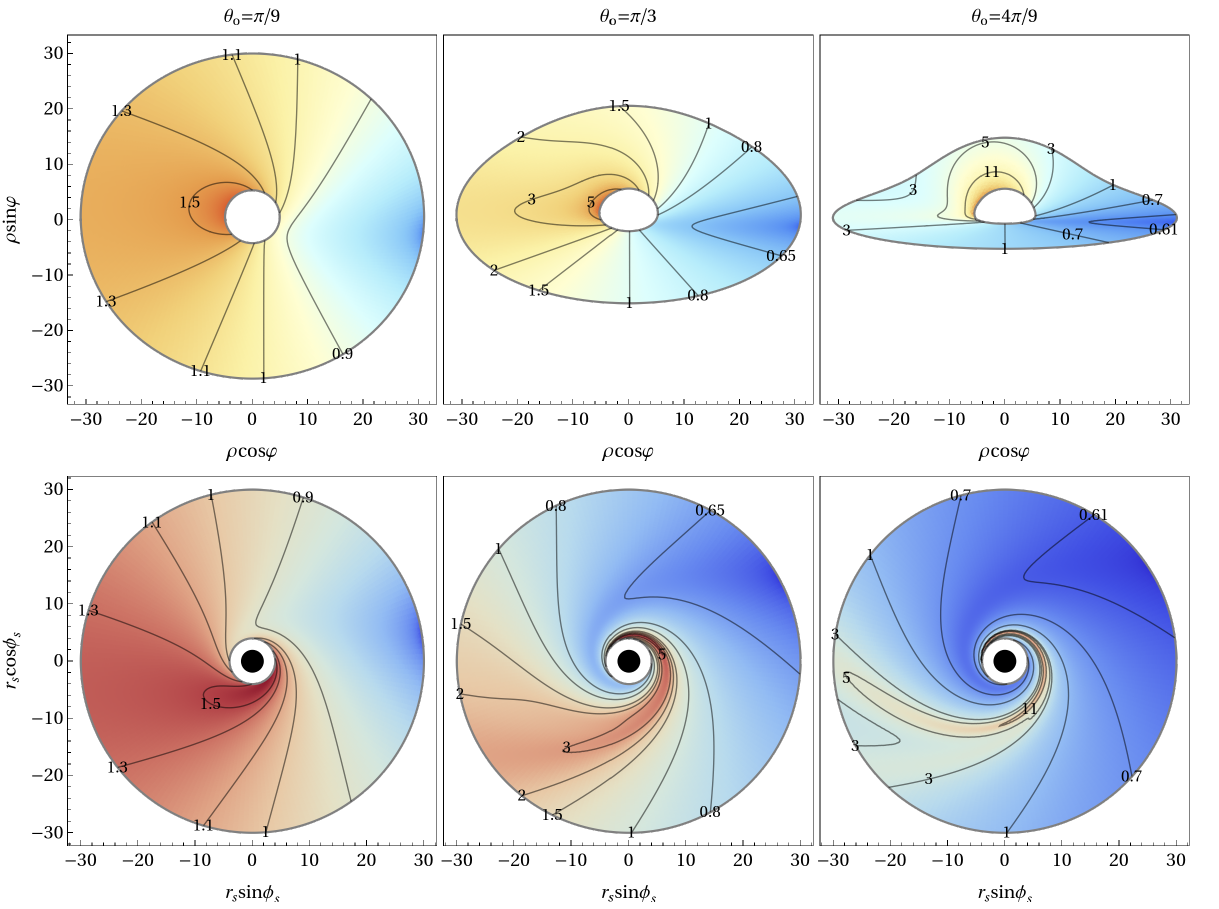}
\caption{Relative magnification factor of corotating point sources distributed on the accretion disk on image plane (top panels) and on source plane (bottom panels) for selected inclination angles. Here, we set the angular speed to be $\omega^{(i)}=1/(2 r_\text{s,0}^{(i)})$.  \label{F6}} 
\end{figure}

We also present the magnification factor across the accretion disks with the angular speed, $r_{\rm s,0}^{(i)}\omega^{(i)}=0.5$, in Fig.~\ref{F6}. By comparing with Fig.~\ref{F5}, it is demonstrated that the rotation speed can modulated the magnification factor across the accretion disks.   

\section{Enhancement of the radiation flux due to the slow-light effect\label{V}}

\begin{figure}
	\includegraphics[width=0.7\linewidth]{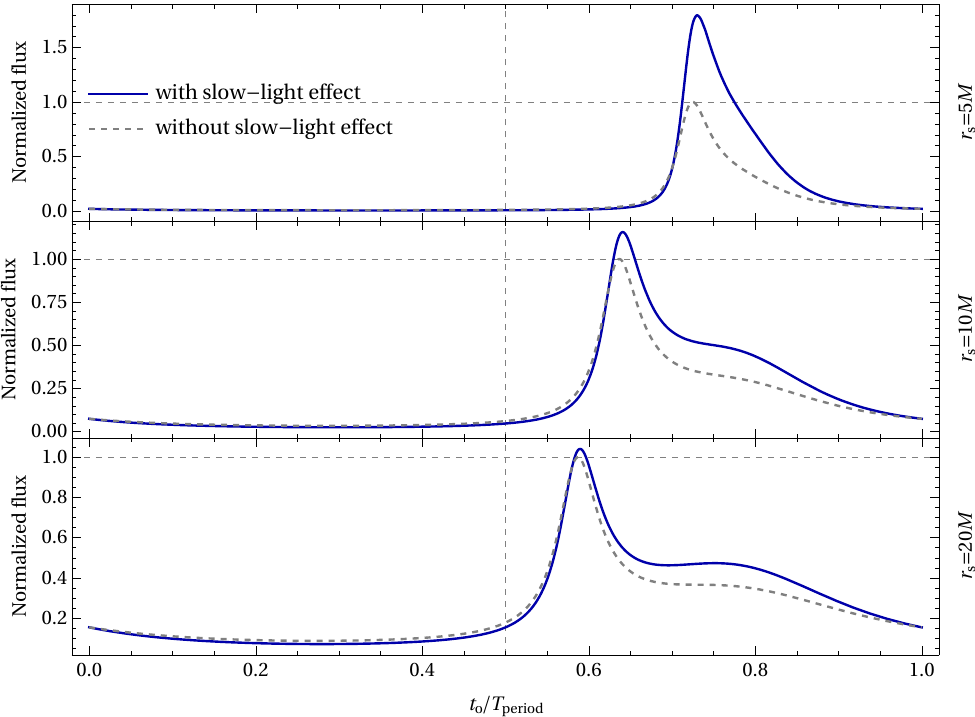}
\caption{Light curves of corotating hotspots on Keplerian disks with or without the slow-light effect. Here, we set the hotspots situated at different distance to the black hole. \label{F7}} 
\end{figure}

As illustrated in Eq.~(\ref{1}), one of the observational signatures of the corotating hotspot is the temporal radiation flux, known as the light curve. For the hotspot localized to a small portion of the accretion disks, its radiation flux with or without considering slow-light effect can be rewritten as
\begin{subequations}
	\begin{eqnarray}  \label{14}
	 F_\text{slow-light}(t_\text{o}) &=& g^3I_\text{emt}\mathcal{D}_\text{o}\mathcal{P}\mu_\text{mov}\Delta \sigma~, \label{14a}\\ F(t_\text{o}) &=& g^3 I_\text{emt} \mathcal{D}_\text{o} \mathcal{P} \mu \Delta \sigma~. \label{14b}
\end{eqnarray}
\end{subequations}
The quantities $\mathcal{D}_\text{o}$, $\mathcal{P}$ and $\Delta \sigma$ reduce to constants for distant observer and hotspot in Keplerian motion (i.e., the hotspot on a circular orbit of fixed radius $r_\text{s}$). Ideally, we treat the hotspot as a point-like source $\Delta \sigma \rightarrow 0$. Here, the impact of redshift factor $g$, including gravitational redshift and Doppler redshift, can not be neglected since the source is moving rapidly in the strong-field regime.
In this study, we focus on the relativistic effect, thereby adopting a uniform intrinsic specific intensity $I_\text{emt}$. 
In Fig.~\ref{F7}, we present the light curves of a corotating hotspot on Keplerian disk with or without the slow-light effect. 
In the absence of the slow-light effect, the magnification factor is given by Eq.~(\ref{6}), while in the presence of the slow-light effect, it is given by Eq.~(\ref{11}). It is found that the radiation flux is enhanced due to the slow-light effect, especially at the peaks of the light curve profiles. This enhancement becomes more significant for the corotating hotspot located closer to the black hole.

\begin{figure}
	\includegraphics[width=1\linewidth]{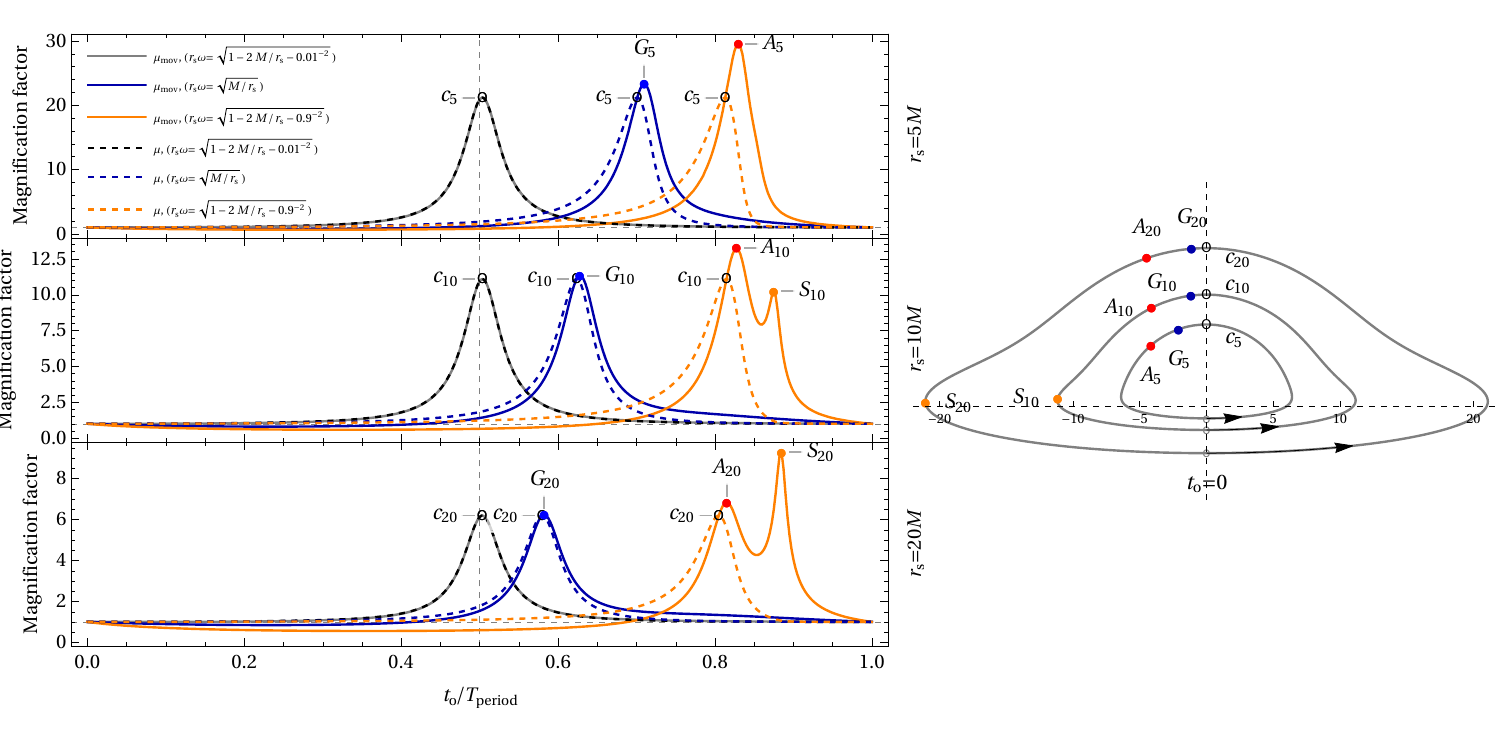}
\caption{\textit{Left panel}: The temporal profiles of magnification factor for hotspot in different angular speeds, with or without the slow-light effect. Besides, the hotspot corotating with Keplerian disk with $ \omega = \sqrt{M/r_\text{s}^3}$, we consider the hotspot corotating with a constant speed $v[=f_\text{s}^{-1/2}r_\text{s}\omega]$, namely, $\omega=r_\text{s}^{-1}\sqrt{1-2M/r_\text{s}-v^{-2}}$. The peaks of magnification factor profiles are denoted as $c_{r_\text{s}}$ (peaks consistent with \textbf{c}onventional results), $G_{r_\text{s}}$ (peaks from circular \textbf{G}eodesic motion), $A_{r_\text{s}}$ (peaks from \textbf{A}ccelerating motion) and $S_{r_\text{s}}$ (peaks arising from \textbf{S}low-light effect). \textit{Right panel}: corresponding positions of these peaks on image plane. Here, the hotspots are set to be rotating anticlockwisely around the black hole. \label{F8}} 
\end{figure}

Notably, the specific fluxes in Eqs.~(\ref{14}) are proportional to $g^3\mu$, collectively determined by the redshift factor and the relative magnification factor. Because the redshift factor of a hotspot is unaffected whether or not the slow-light effect is included, the differences in the light curves should be attributed to the magnification factor. To illustrate this, we present the temporal profile of magnification factor in Fig.~\ref{F8}. 
Owing to the slow-light effect, the peaks of the magnification factor profiles are enhanced and shifted, as shown by comparing the solid and dashed blue curves in the right panel of Fig.~\ref{F8}. To study the influence of disk's rotation speed on the magnification factor, we also consider the corotating hotspots with speed $r_\text{s}\omega/\sqrt{f(r_\text{s})}=0.01$ and $0.9$. It shows that the slow-light effect becomes more significant for the rapidly corotating hotspots, as indicated by the orange curves in Fig.~\ref{F8}. In the right-middle and right-bottom panels of Fig.~\ref{F8}, which correspond to distant sources, the secondary  peaks emerge, denoted by $S_\text{10}$ and $S_\text{20}$. The left panel of Fig.~\ref{F8} shows the image positions of the hotspots when they reach their maxima of magnification factor profiles. Without the slow-light effect, the hotspots at the maxima of the magnification factor are located  behind the black hole as seen by a distant observer, which is consistent with the results in Fig.~\ref{F3}. With the slow-light effect, the corotating hotspots at the maxima are located at positions which are shifted relative to the image positions behind the black hole. This is consistent with the results in Fig.~\ref{F5}. 

For all these cases, we perform a consistency checking numerically, 
\begin{eqnarray}
	\langle \mu \rangle = \langle \mu_\text{mov}\rangle~, \label{15a}
\end{eqnarray}
where the time average of quantity $A$ is defined as
\begin{eqnarray}
	\langle A \rangle \equiv \frac{1}{T_\text{period}}\int_0^{T_\text{period}} A \mathrm{d}t~.
\end{eqnarray}
This indicates that the time-averaged magnification factor is unaffected by the slow-light effect. As illustrated in the left-bottom panel of Fig.~\ref{F4}, some moving sources are magnified while others are demagnified relative to the sources on static disk. Namely, Eq.~(\ref{15a}) holds because the total image size of the accretion disk remains unaffected by velocity of the corotating accretion flow. 

\begin{figure}
	\includegraphics[width=0.75\linewidth]{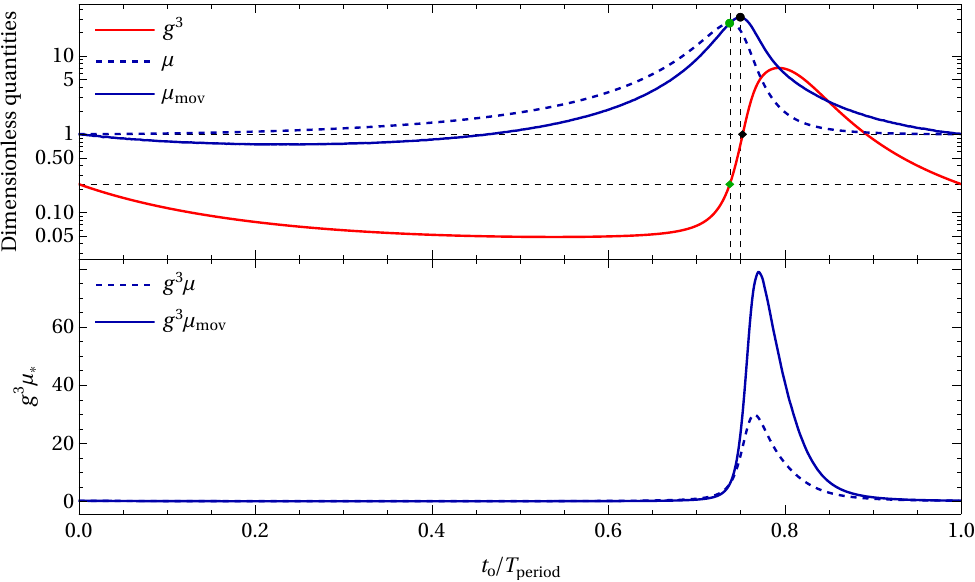}
	\caption{\textit{Top panel}: redshift factor $g$, magnification factor $\mu$ [Eq.~(\ref{6}) without the slow-light effect] and magnification factor $\mu_\text{mov}$ [Eq.~(\ref{11}) with the slow-light effect] as function of observed time. \textit{Bottom panel}: $g^3\mu$ or $g^3\mu_\text{mov}$ as function of observed time $t_\text{o}$. We model the hotspot corotating with the Keplerian disks, $\omega=\sqrt{M/r_\text{s}^3}$. In this setup, the radiation flux is proportional to $g^3 \mu_\ast$. The inclination angle of the disk is set to be $4\pi/9$, and the orbital radius is set to be $r_\text{s}=4M$. \label{F9}}
\end{figure}

Interestingly, the magnification factor is only slightly  enhanced  by the slow-light effect, as shown in the left panel of Fig.~\ref{F8}, while the radiation flux is enhanced significantly, as presented in Fig.~\ref{F7}. This flux enhancement should be interpreted in association with the redshift factor. We illustrate the physical mechanism causing this enhancement in Fig.~\ref{F9}. The radiation flux, proportional to $g^3 \mu$, will be significantly enhanced if the peaks of $g$ and $\mu$ tend to align. In this sense, as shown in the bottom panel of Fig.~\ref{F9}, the slow-light effect results in the peak of $\mu$ shifting towards alignment with the peak of $g$, serving as the main origin of the flux enhancement. Specifically, for the source near the black hole $r_\text{s}=4M$ shown in Fig.~\ref{F9}, the flux maximum is enhanced by a factor of three relative to the case neglecting the slow-light effect.

\begin{figure} 
	\includegraphics[width=0.7\linewidth]{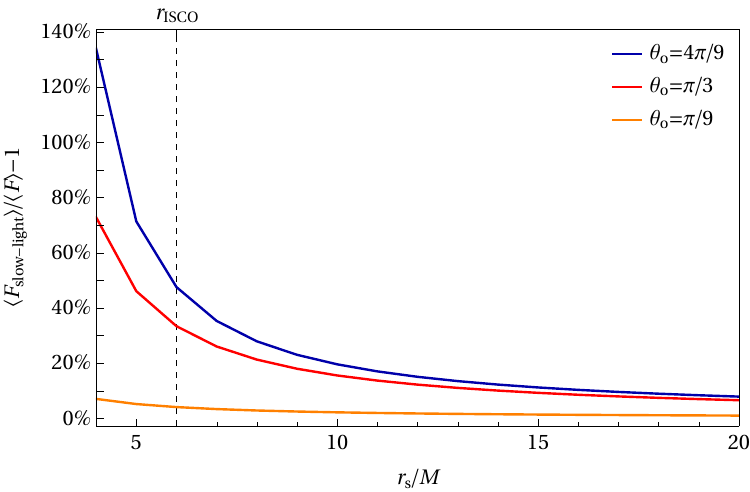}
	\caption{The enhancement ratio for the time-averaged radiation fluxes as function of radial location $r_\text{s}$ of Keplerian disks for selected disk inclination angles. \label{F10}}
\end{figure}

In contrast to the invariance of the time-averaged magnification factor in Eq.~(\ref{15a}), the time-averaged flux is notably altered by the slow-light effect. We quantify the flux enhancement by defining a ratio as follows,
\begin{eqnarray}
	\langle F_\text{slow-light} \rangle  / \langle F \rangle = \frac{\displaystyle\int_0^{T_\text{period}}g^3 \mu_\text{mov} \mathrm{d}t}{\displaystyle \int_0^{T_\text{period}}g^3 \mu \mathrm{d}t
	}~. \label{17}
\end{eqnarray}
For fixed hotspot's orbital radius $r_\text{s}$, the observer's location, and the constant $I_\text{emt}$, the quantities $\mathcal{D}_\text{o}$, $\mathcal{P}$ and $\Delta\sigma$ are independent of time. Consequently, these quantities cancel out in Eq.~(\ref{17}) as they are the same value in both the numerator and the denominator.
In Fig.~\ref{F10}, we present the enhancement ratio $\langle F_\text{slow-light} \rangle  / \langle F \rangle$ for given radial locations $r_\text{s}$ of the disk. This enhancement becomes significant with increasing inclination angle or decreasing orbital radius. 
In the standard model of the geometrically thin accretion disk (the Novikov-Thorne disk \cite{1973blho.conf..343N}), the inner boundary is determined by the innermost stable circular orbit at $r_\text{ISCO}=6M$ for Schwarzschild black hole. Our results, as illustrated in Fig.~\ref{F10}, reveal that the slow-light effect has non-negligible contributions to the emission from the ISCO region. Specifically, the slow-light effect induces flux enhancement of approximately 50\% for inclination $\theta_{\mathrm{o}}=4\pi/9$ and emission originating at $r_{\mathrm{ISCO}}$.
We demonstrate that the flux enhancement induced by the slow-light effect is non-negligible for emission on the accretion disk in the strong-field regime.

\section{Conclusions and discussions\label{VI}}

In this paper, we investigated the impact of the slow-light effect on the observational signatures of hotspots corotating with Keplerian disks.  By implementing the slow-light effect through a time-series reconstruction, we demonstrate that the magnification factor is modulated by the orbital velocity of the hotspot. Specifically, the corotating hotspots at the maxima of magnification are located at positions shifted relative to the image positions   behind the black hole. This causes the peak of the magnification to shift towards the peak of the redshift factor, ultimately leading to flux enhancement. This enhancement becomes particularly pronounced for corotating hotspots in the strong-field regime, indicating that the slow-light effect is essential for accurately modeling high-energy emission in the vicinity of black holes.

Previous studies \cite{1992A257594B,1994ApJ...425...63B,Zhu:2024vxw}, as reviewed in Ref.~\cite{Bozza:2010xqn}, evaluated the quantity $\Sigma$ [Eq.~(\ref{8})] under the fast-light approximation. The main objective of this study is to extend the slow-light treatment to the quantity $\Sigma$. To explicitly isolate the impact of the slow-light effect, we introduce the magnification factor $\mu$ by decomposing $\Sigma$. This impact is demonstrated in the light curves [Eqs.~(\ref{14})], where we compare the results obtained the magnification factor using the slow-light treatment $\mu_\text{mov}$ [Eq.~(\ref{11})] with that from the fast-light approximation $\mu$ [Eq.~(\ref{8})].

Notably, previous studies have concluded that the slow-light effect modifies the flux of accretion disks by no more than 10\% relative to the fast-light approximation 
\cite{Dexter:2010pe,Bronzwaer:2018lde,Moscibrodzka:2021fxv,White:2022paq}. However, the flux enhancement found in this study is more significant than expected, as shown in Fig.~\ref{F10}.
The discrepancy arises because previous studies validate the robustness of the fast-light approximation for regimes with a modest optical depth, which masks the differences between the results in slow-light and fast-light treatments. 
If we focus on the infrared flares \cite{GRAVITY:2018sef,GRAVITY:2023avo}, where the environment around the black hole is optically thin, our result could be valid. In this regime, the flux  can be significantly amplified by the redshift factor, as illustrated in Fig.~\ref{F9}.


This study employs the toy model of the hotspots \cite{1992A257594B,1994ApJ...425...63B}, treating both the specific intensity $I_\text{emt}$ and the emission region $\Delta \sigma$ as constants. In a realistic scenario, $I_\text{emt}$ depends on the specific emission mechanism, and the finite emission region $\Delta \sigma$ can be elongated by the shear induced by the disk's differential rotation \cite{Levis:2023tpb}. 
These physical effects also can be straightforwardly incorporated into our  framework, enabling a comparison of the slow-light and fast-light treatments for a more realistic disk.

\smallskip
{\it Acknowledgments}:  this work is supported by the National Natural Science Foundation of China under grants No.~12305073 and No.~12347101. The author thanks Prof. Xin Li for useful discussions, Prof.~Valerio Bozza for helpful comments, and the anonymous referee for their valuable suggestions.

\bibliography{ref}

\appendix

\section{Projection coefficient $\mathcal{P}$\label{A}}

The expression projection coefficient $\mathcal{P}$ is not be easily obtained via a mathematical derivation and is given empirically. In this part, we will show that there is an numerical scenario for obtaining projection coefficient, which does not completely satisfy assumption ii). 

Based on assumption i), the relative magnification factor is proportional to
\begin{eqnarray}
	\mu &\propto&   \frac{\xi}{\sqrt{h}} \left| \frac{\partial (\xi, \varphi)}{\partial (r_\text{s}, \phi_\text{s})}
	\right|=:P~, 
\end{eqnarray}
If considering the magnification factor relative to the front side of the disk, the projection coefficient can be given by
\begin{eqnarray}
	\mathcal{P}_\text{disk}= \lim_{\varphi\rightarrow\pi/2} P \label{A2}
\end{eqnarray}
In Fig.~\ref{F6}, we present the comparison between the numerical projection coefficient in Eq.~(\ref{A2}) and the that used in our main context, i.e., $\mathcal{P}_\text{eff}=\sqrt{f_\text{o}}|\cos\theta_\text{o}|$. In the large inclination angle $\theta_\text{s}\rightarrow \pi/2$, the definition of the projection coefficient $\mathcal{P}_\text{disk}$ in Eq.~(\ref{A2}) is consistent with the $\mathcal{P}_\text{eff}$. This validates the analytical expression for the projection coefficient $\mathcal{P}_\text{eff}$.
\begin{figure}[t]
	\includegraphics[width=1\linewidth]{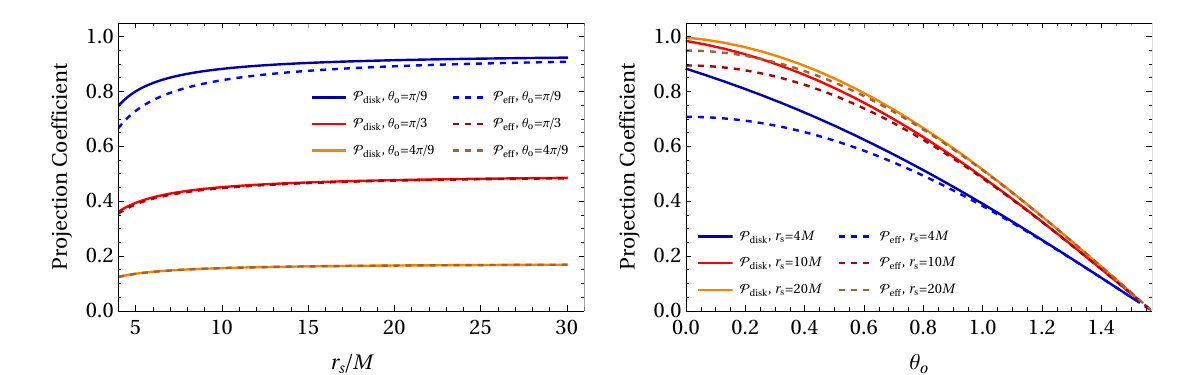}
\caption{A comparison of the projection coefficients, $\mathcal{P}_\text{disk}$, in Eq.~(\ref{A2}) and that in the form of $\mathcal{P}_\text{eff}= \sqrt{f_\text{s}}|\cos\theta_\text{o}|$. \label{F11}}
\end{figure}

It is important to clarify that the flux enhancement ratio in Eq.~(\ref{17}) is unaffected by the choice of projection coefficient. Namely, based on Eq.~(\ref{17}) and $\mu\mathcal{P}_\text{eff} = \mu'\mathcal{P}_\text{disk}$, we have
\begin{eqnarray}
	\langle F_\text{slow-light} \rangle  / \langle F \rangle = \frac{\frac{\mathcal{P}_\text{eff}}{\mathcal{P}_\text{disk}} \int_0^{T_\text{period}}g^3 \mu_\text{mov} \mathrm{d}t}{ \frac{\mathcal{P}_\text{eff}}{\mathcal{P}_\text{disk}}\int_0^{T_\text{period}}g^3 \mu \mathrm{d}t
	} = \frac{\int_0^{T_\text{period}}g^3 \mu'_\text{mov} \mathrm{d}t}{ \int_0^{T_\text{period}}g^3 \mu' \mathrm{d}t
	} ~. 
\end{eqnarray}

\section{Consistency between numerical and analytical methods}
\begin{figure*}
	\centering
	\begin{subfigure}[t]{0.55\textwidth}
	\includegraphics[width=1\linewidth]{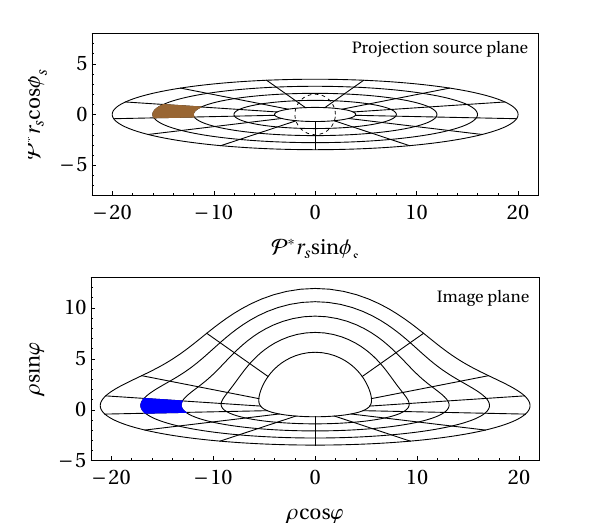}
		\end{subfigure} 
	 \begin{subfigure}[t]{0.44\textwidth}  
		\includegraphics[width=1\linewidth]{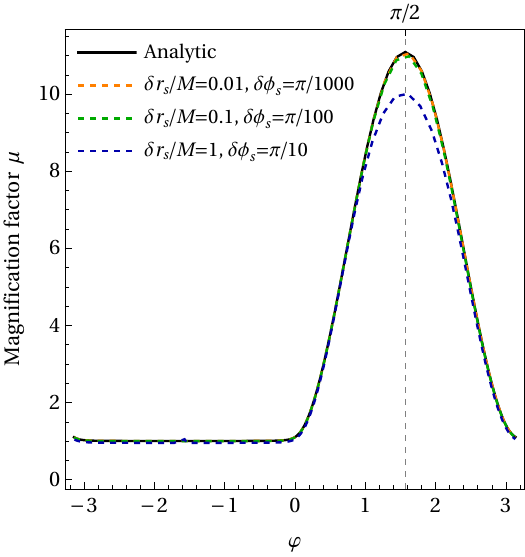}
			\end{subfigure} 
			\caption{\textit{Top-left panel}: the source, denoted as brown region, distribution on the thin accretion disk presented on the projected source plane. \textit{Bottom-left panel}: corresponding images of the source, denoted as blue region, distribution on the thin accretion disk presented on image plane. Here, we set the center of the source region located at $r_\text{s}=14M$, accretion disk at the inclination angle $\theta_\text{o}=4\pi/9$ and observers' distance far away from the black hole, $r_\text{o}=10^5M$. \textit{Right panel}: the magnification factor as function of $\varphi$, the polar coordinate of image plane, for different size of the sources. The $\delta r_\text{s}$ and $\delta \phi_\text{s}$ describe the radial and azimuthal extension of the sources, respectively.  \label{F2}}
\end{figure*}

The results in Fig.~\ref{F2} validate the magnification factor presented in Eq.~(\ref{8}). Here, the numerical method for magnification factor is implemented via separately calculating the source size on the projected source plane and the corresponding image size on the image plane, as illustrated in the left panel of Fig.~\ref{F2}. The finite size effects of the sources inevitably introduce deviations in the magnification factor, as shown in the right panel of Fig.~\ref{F2}. The $\mu$ reduces to the analytical result in Eq.~(\ref{8}) as the source shrinks to a point-like source. Notably, for sources located in front of the black hole relative to the observer (i.e., $\varphi \in [0,\pi]$), we have $\mu \simeq 1$, satisfying assumption ii). 

\end{document}